\documentclass[english,prl,aps,twocolumn,floatfix]{revtex4-1}

\usepackage[linktocpage,bookmarksopen,bookmarksnumbered]{hyperref}
\usepackage{graphicx}
\usepackage{dcolumn}
\usepackage{amsmath,graphics,epsfig,color,verbatim,ulem}
\usepackage{amssymb}
\usepackage{babel}

\begin{document}

\title{Total energy calculations using DFT+DMFT: computing the pressure phase diagram of the rare earth nickelates}

\author{Hyowon Park,$^{1,2}$ Andrew J. Millis,$^{2}$ and Chris A. Marianetti$^{1}$}

\affiliation{$^{1}$Department of Applied Physics and Applied Mathematics, Columbia University, New York, New York 10027, USA \\
             $^{2}$Department of Physics, Columbia University, New York, New York 10027, USA}

\date{\today}

\begin{abstract}
A full implementation of the $ab$ $initio$ density functional plus dynamical mean field theory (DFT+DMFT) formalism to perform total energy calculations and structural relaxations  is proposed and implemented.  The method is applied to the structural and metal-insulator transitions of the rare earth nickelate perovskites as a function of rare earth ion,  pressure, and temperature. In contrast to previous DFT and DFT+$U$ theories, the present method accounts for the experimentally observed structure of $La$NiO$_3$ and  the insulating nature of the other perovskites, and quantitatively reproduces the metal-insulator and structural phase diagram in the plane of pressure and rare earth element. The temperature dependence of the energetics of the phase transformation indicates that the thermal transition is driven by phonon entropy effects.

\end{abstract}

\maketitle

\section{Introduction}

Understanding the interplay of the  quantum mechanics of strongly interacting electrons and the crystal structure of real materials is a fundamental challenge for modern materials theory. In correlated electron materials, metal-insulator transitions and other important electronic phenomena often occur in conjunction with large amplitude lattice distortions and changes in crystal symmetry and theoretical methods must handle both on the same footing. 
The combination of density functional theory (DFT) and dynamical mean field theory (DMFT)~\cite{Kotliar:06} enables calculations of many-body physics in the context of a realistic crystal structure.  Most applications of the method to date have featured the computation of spectroscopic properties in a fixed structure determined either through experiment or DFT or DFT+$U$. The DFT+DMFT method has not been widely used to compute which is the favored structure. In this paper we show that total energy calculations within the DFT+DMFT formalism correctly reproduced the nontrivial coupling between the structural and metal-insulator transitions in the strongly correlated rare earth nickelate materials. 
%In the absence of information about energetics in the context of a realistic description of the electronic structure, phase diagrams and lattice structures cannot be predicted.  

Total energy calculations have been implemented in the DMFT framework with
varying degrees of sophistication.  In pioneering work, the energy of the
$\delta$-Pu was computed as a function of volume~\cite{Savrasov:01}, though an
approximate DMFT impurity solver was used. 
% and the calculated phonon spectra of Pu~\cite{Dai:03}. Another example of the DFT+DMFT energy calculation in $f$ electron system is
The volume collapse transition of  paramagnetic
Ce~\cite{Held:01,McMahan:03,Amadon:06} has also been studied, but full charge self-consistency
was not attempted. 
%More recently
%work on the energetics of transition metal systems with different structures
%has appeared~\cite{Leonov:08,Marco:09,Leonov:10,Leonov:11}.  
More recently, studies have been performed on the energetics of transition metal
systems using full charge self-consistency with approximate DMFT impurity
solvers~\cite{Marco:09}, while other studies have used the Hirsch-Fye quantum
Monte Carlo (QMC) for the DMFT impurity problem but do not include full charge
self-consistency~\cite{Leonov:08,Leonov:10,Leonov:11}. Fully charge
self-consistent calculations using approximate DMFT impurity solvers also have been
performed to study the elastic properties of
Ce~\cite{Amadon:12}, Ce$_2$O$_3$~\cite{Pourovskii2007235101,Amadon:12}, and
Pu$_2$O$_3$~\cite{Amadon:12}.  Very recently, fully charge self-consistent
calculations which use continuous-time QMC to solve the DMFT impurity problem
have been used to calculate the $z$ position of the As atom in iron
pnictides~\cite{Aichhorn:11,Lee:12}  and the thermodynamics of
V$_2$O$_3$~\cite{Lechermann:12}, Ce~\cite{Amadon:13}. These most advanced studies have not yet addressed a phase
transition between two different structures.

In this paper we investigate the structural and metal-insulator phase boundaries of the family of rare earth nickelate perovskites $R$NiO$_3$ as a function of rare earth ion $R$ and pressure.  The rare earth nickelates provide a crucial challenge to theoretical methodologies because they exhibit a  metal-insulator transition which is closely tied to a large-amplitude two-sublattice bond-length disproportionation in which the mean Ni-O bond length becomes larger for Ni sites on one sublattice and smaller on the other~\cite{Medarde:97}. The electronic state has been the subject of substantial discussion~\cite{Alonso:99,Alonso:01,Staub:02,Mazin:07,Medarde:09,Lee:11} but has now been identified as a  site-selective Mott transition~\cite{Park:12}. The location of the phase boundaries in the pressure-temperature plane varies across the rare earth series~\cite{Medarde:97,Alonso:01,Lengsdorf:04,Amboage:04,Amboage:05,Cheng:10}, with $Lu$ having the highest critical temperature and pressure and $La$ remaining metallic down to lowest temperature at ambient pressure.  Standard DFT and DFT+$U$ methods fail to describe the phase diagram, with DFT predicting that all compounds remain metallic and un-disproportionated and DFT+$U$ predicting that all compounds are disproportionated at ambient pressure.   These results establish that strong electronic correlations are crucial to structural phase stability and methods beyond DFT and DFT+$U$ are  required to properly describe them. Here we show that DFT+DMFT succeeds in providing a unified description of the entire class of the rare-earth nickelates using as input only the nature of the atoms, with the interaction parameters ($U$,$J$, double counting correction) fixed for the entire series. 

\section{Total Energy Methodology}

We perform  total energy calculations within the charge-self-consistent DFT+DMFT framework \cite{Kotliar:06}. The total energy is obtained  from  the DFT+DMFT functional $\Gamma[\rho,G]$ using the converged charge density $\rho$ and  local Green's function $G$ as 
\begin{equation}
E^{tot}[\rho,G] = E^{DFT}[\rho]+E^{band}[\rho,G]+E^{pot}[G]-E^{dc}
\label{eq:energy}
\end{equation}
where $E^{DFT}$ is the DFT energy, $E^{band}$ is the Kohn-Sham band energy correction due to the DMFT density matrix and $E^{pot}$ is the DMFT potential energy obtained by the trace of $\Sigma G$. The DFT+DMFT procedure requires values for the on-site interactions and, if standard approximations to the exchange correlation functional are used, a `double counting correction' ~\cite{Anisimov91,Sawatzky:94,Amadon08,Karolak10} formally written here as $E^{dc}$. %A key limitation of the DFT+DMFT method is the  double counting term, which is a necessity if standard approximations to the exchange-correlation functional like LDA or GGA are to be used. 
Typically, the double-counting energy is expressed as a function of the total occupation, $N_d$, of the correlated subspace, and the result of the DMFT and charge self consistency is to fix $N_d$ and thus the mean $p-d$ energy splitting at particular values.  Previous work has shown that the results are very sensitive to the resulting value of $N_d$ and hence to the form of  the double-counting~\cite{Wang:12,Park:12,Dang:13}. Here we use the functional form of  the fully-localized-limit $E^{dc}$ ~\cite{Sawatzky:94},
but allow for a  prefactor $U^\prime$ which may be different from $U$:
\begin{eqnarray}
E^{dc} &=& \frac{U'}{2}N_d(N_d-1)-\frac{5J}{4}N_d(N_d-2),%\\
%\hat{V}^{dc} &=& U'(N_d-\frac{1}{2})-\frac{5J}{2}(N_d-1)
\label{eq:DC}
\end{eqnarray}
with the double counting potential $V^{dc}=\partial E^{dc}/\partial N_d$. 
%In this work we set  $U^\prime$=4.8eV($<U$=5eV), only slightly different from the conventional  choice of $U^\prime=U$. This small difference of $U^\prime$ from $U$ leads to a small change in  $N_d$ which we show  is necessary for the proper description of the entire series of materials. 
In this work we choose a $U'$ value (4.8eV) only slightly different from the
conventional choice of $U'=U$ (5.0eV) and
we fix this $U'$ for all $R$NiO$_3$ series used in this paper. Compared to the conventional choice of $U'=U$, our choice has the effect of modestly increasing the energy splitting between oxygen $p$ and Ni $d$ states and modestly decreasing the $d$ occupancy.
As we will see, this choice of  $U'$  provides a correct and consistent account of the physics of rare earth nickelates across the entire phase diagram, providing strong a posterior evidence in favor of this ansatz.
Not only the energetics, but also the DFT+DMFT spectral function of nickelates computed using the reduced $U'$ value
is more consistent with experimental spectra than is the conventional choice~\cite{supple}.
We further observe that the conventional choice of double counting has no clear theoretical justification~\cite{Karolak10} and fails to produce
the correct structural properties in nickelates (see Fig.$\:$\ref{fig:ecurve}) while in early transition metal oxides it fails to reproduce the known Mott insulating behavior~\cite{Dang:13}.

Our calculations use the Vienna ab-initio simulation package~\cite{Kresse19991758,Kresse199611169} (VASP) with the Perdue-Burke-Ernzerhof exchange-correlation functional and a $k$-point mesh of $6\times6\times6$  for the $Pbnm$ and $P2_1/n$ structures and and  $8\times8\times8$ mesh for the $La$NiO$_3$ $R\bar{3}c$ structure
with an energy cutoff of 600eV. The Kohn-Sham Hamiltonian is represented using maximally localized Wannier function (MLWF)~\cite{Wannier} defined in an energy window including the full $p-d$ manifold, in order to allow for a tractable DFT+DMFT implementation in a plane-wave basis. The correlated subspace is defined to be the Wannier orbitals corresponding to atomic-like Ni-$d$ orbitals defined from the Wannier construction.  They interact via the fully rotationally invariant Slater-Kanamori interactions. The intra-orbital Coulomb interaction  $U$ is set to 5eV and the  Hund coupling $J$ to 1eV for all calculations reported here. We do not include spin-orbit coupling. The filled and electronically inert $t_{2g}$ orbitals  are treated by a Hartree-Fock approximation while the $e_g$ orbitals are treated by DMFT. The impurity model is solved using the hybridization expansion version of the  numerically exact continuous time QMC method~\cite{Werner:06,Haule:07}; temperatures as low as 0.01eV are accessible. Computation of atomic forces have not yet been implemented in the MLWF basis. %saying due to technical challenges is the same as saying nothing so I left it out due to technical challenges. 
Therefore, at a specified volume we minimize the DFT+DMFT total  energy along 
a one-dimensional path that interpolates between the $Pbnm$ and highly distorted $P2_1/n$ structures, 
and to determine the global minimum we then find the unit cell volume at which the total energy is minimized. 
To construct the one dimensional
path at fixed volume we determine the internal cell coordinates
and cell shape for the $Pbnm$ structure by relaxing using DFT
and for the $P2_1/n$ structure by relaxing using DFT+$U$.
Interpolating between the two structures 
defines a one dimensional path for this specified volume. 
We parametrize the path by the value of the difference $\delta a$ between the mean Ni-O bond lengths 
in the two sublattices in the $P2_1/n$ structure. This prescription is chosen because $Pbnm$ 
is unstable within DFT+$U$ while $P2_1/n$ is unstable within DFT.  
It should be noted that this algorithm is well defined and provides 
a reasonable approximation to the global minimization. Also, for a DFT+DMFT path, 
the fixed volume in Fig.$\:$\ref{fig:ecurve} is slightly larger than the volume 
at a global minimization since the theoretical pressure can not be computed.

\section{Results: Phase Diagram}

\begin{figure}[!htbp]
\includegraphics[scale=0.45]{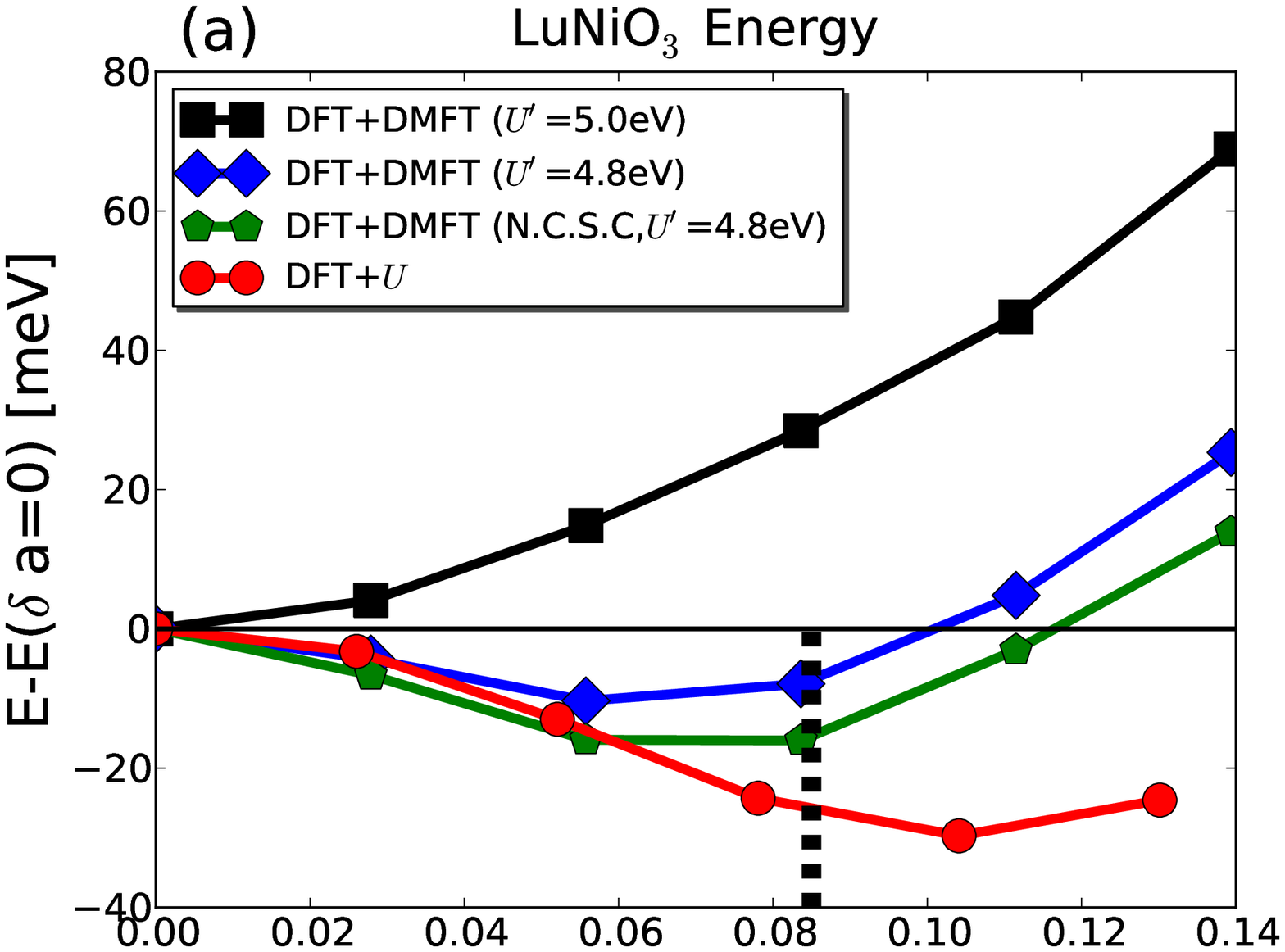}
\includegraphics[scale=0.45]{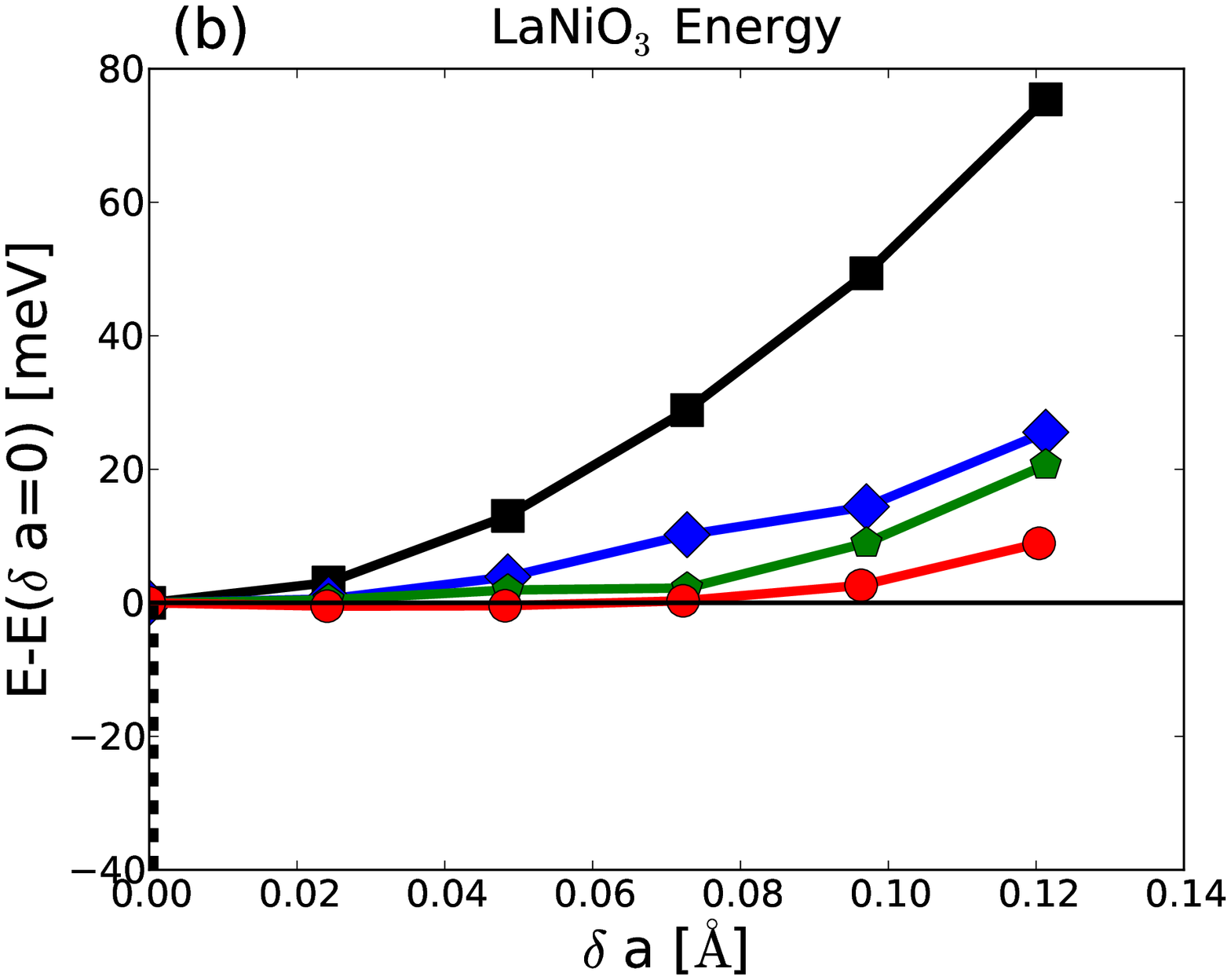}
\caption{(Color online)
Total energy as a function of bond-length difference $\delta a$  for $Lu$NiO$_3$ (upper panel)  and $La$NiO$_3$ (lower panel) calculated as described in the text using fully charge self consistent DFT+DMFT with original (squares, black online) and modified (diamonds, blue online) double counting correction and compared to DFT+DMFT energies computed using the DFT charge density (N.C.S.C, pentagons, green online) and to energies obtained from the DFT+$U$ method (circles, red online). The experimentally determined values ($\delta a=0.085\AA$ ($Lu$NiO$_3$) and $\delta a=0$ ($La$NiO$_3$)) are indicated by vertical dashed lines.  The interaction parameters for both DFT+DMFT (T=116K) and DFT+$U$ (implemented in VASP) are $U$=5.0eV and $J$=1.0eV. 
\label{fig:ecurve}}
\end{figure}

We begin by presenting in Fig.$\:$\ref{fig:ecurve} the energy as a function of distortion for the two end members of the phase diagram: $La$NiO$_3$ and $Lu$NiO$_3$. Comparison of the diamonds (blue online) and pentagons (green online) shows that the full charge self-consistency has only a small effect on the total energy. Given that full charge  self-consistency is extremely costly from a computational standpoint, all remaining calculations are performed using the converged DFT charge density.  The DFT+DMFT total energy curve obtained with the modified double counting $U'$=4.8eV  are in a good agreement with the experimental values.  $La$NiO$_3$ is correctly found not to disproportionate while  $Lu$NiO$_3$ is found to disproportionate and minimizing the energy yields a nearly correct  Ni-O bond-length difference and an insulating ground state.  
If the standard double counting is used $Lu$NiO$_3$ is wrongly predicted not to disproportionate. 
Finally we note that the DFT+$U$ approximation strongly overestimates the amplitude of the disproportionation in $Lu$NiO$_3$ and wrongly predicts that $La$NiO$_3$ is disproportionated (although the energy minimum is very shallow).

\begin{figure}[!htbp]
\includegraphics[scale=0.45]{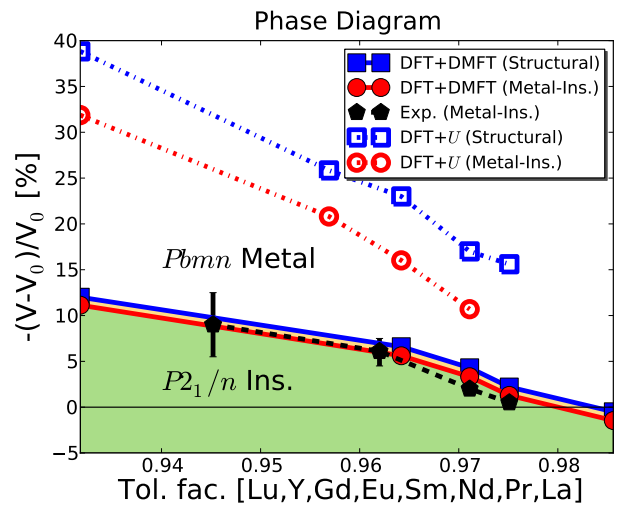}
\caption{(Color online)
Metal-insulator and structural phase diagram computed using DFT+DMFT (solid symbols and solid lines) as a function of unit cell volume for the series of rare earth ions and compared to results of DFT+$U$ calculations (open symbols, dashed lines) and to  experimental data (pentagons and dashed lines, black on-line) obtained for ($Y$,$Eu$,$Nd$,$Pr$)NiO$_3$ using the data Ref.$\:$\onlinecite{Cheng:10} as explained in the text. $V_0$ is determined as the calculated ambient pressure equilibrium volume for each material while the tolerance factor is determined from the distances $d_{R-O}$ and $d_{Ni-O}$ as $d_{R-O}/d_{Ni-O}\sqrt{2}$~\cite{Medarde:97}.
The parameters for the  DMFT calculations are $T$=116K, $U$=5eV, $U'$=4.8eV and $J$=1eV.
\label{fig:phasedmft}}
\end{figure}

Having established our approach for the end members at ambient pressure and low temperatures, we now compute the phase diagram as a function of unit cell volume (tuned experimentally by pressure). Fig.$\:$\ref{fig:phasedmft} shows the calculated DFT+DMFT phase boundaries for the structural (squares and  solid lines, blue online) and the metal-insulator (circles and solid lines, red online)  transitions compared to the results of DFT+$U$ calculations (open symbols, blue and red online)  and to experiment (pentagons and dashed lines, black online). To obtain the experimental results we used the results of Ref.$\:$\onlinecite{Cheng:10} for the pressure-driven metal-insulator transitions in pressure data in $(Y,Eu,Nd,Pr)$NiO$_3$.  Low $T$ ($\sim 100K)$ data were used where available. For $Y$NiO$_3$ and $Eu$NiO$_3$, only higher $T$ data were available and results for $T=100K$ were  extrapolated from the high temperature results using a $T$-dependence derived from the published data on $Nd$ and $Pr$ compounds. The slow variation of the critical temperature $T_{MI}$ with pressure justifies the extrapolation. The error bars indicate the uncertainties arising from the extrapolation. The critical pressure is converted to a reduced volume using the DFT estimate of the pressure-volume curve obtained from DFT calculations.  We note that Ref.$\:$\onlinecite{Mazin:07}  reports results for $Lu$NiO$_3$ which correspond to a transition at a much smaller volume difference (much smaller critical pressure) inconsistent with our calculations or the trends reported in  Ref.$\:$\onlinecite{Cheng:10}. Possible reasons for the discrepancy are discussed in Ref.$\:$\onlinecite{Cheng:10}.

The calculated DFT+DMFT phase diagram is in good agreement with experiment, predicting correctly that at ambient pressure  all rare-earth nickelates are bond-length disproportionated and insulating  except $La$NiO$_3$ and  reproducing quantitatively the critical volume at which the insulating and distorted state is destroyed. A prediction is that under 1.5\% volume expansion $La$NiO$_3$ would  undergo a metal-to-insulator transition. By contrast the DFT+$U$ method strongly overestimates the critical compression needed to destroy the insulating phase.

\begin{figure}[!htbp]
\includegraphics[scale=0.45]{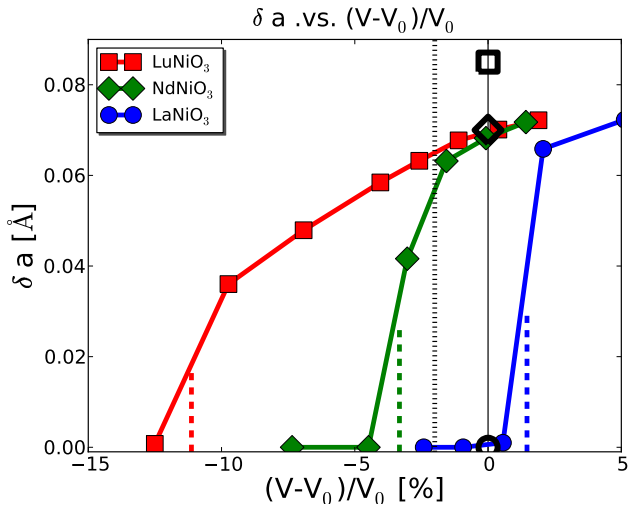}
\caption{(Color online) 
The average Ni-O bond-length difference $\delta a$ determined by minimizing the DFT+DMFT energy for the two inequivalent Ni atoms in $R$NiO$_3$  as a function of a reduced volume for materials indicated. 
%LuNiO$_3$ (red square dots), NdNiO$_3$ (green diamond dots), and LaNiO$_3$ (blue circle dots) results are shown.
%The DFT+DMFT total energy calculation is adopted to determine $\delta a$ at the energy minimum for each fixed volume.
The experimental $\delta a$ values for $Lu$NiO$_3$ (black square empty dot), $Nd$NiO$_3$ (black diamond empty dot), and $La$NiO$_3$ (black circle empty dot) at the respective  equilibrium volumes (x=0) are also shown for comparison.
The theoretically determined critical volumes at which  the metal-insulator transition occurs in each material are  shown as vertical dashed lines. The vertical black dotted line  shows the reduced volume for $Nd$NiO$_3$ at which the experimental metal-insulator transition occurs. 
 \label{fig:bondvsvol}}
\end{figure}

Fig.$\:$\ref{fig:bondvsvol} displays the bond-length disproportionation $\delta a$ of the two inequivalent Ni-O octahedra
obtained at different volumes for $Lu$NiO$_3$ (square, red online), $Nd$NiO$_3$ (diamond, green online), and $La$NiO$_3$ (circle, blue online) along with experimental values obtained at ambient pressure (open symbols). The $\delta a$ values obtained from DFT+DMFT are very close to  the experimental values.
The qualitative features of the $\delta a$ vs volume curves are similar for all nickelates. 
As also seen in Fig.$\:$\ref{fig:phasedmft}, the calculated  insulator to metal transition (marked by  vertical dashed lines) occurs after the onset of the structural distortion. %with the finite but small $\delta a$, and at a slightly increased pressure, the $P2_1/n$ structure undergoes a structural transition to the $Pbnm$ ($\delta a$=0) structure. In LuNiO$_3$, the metal-insulator transition occurs at around 11\% contraction of the volume from the equilibrium one and the structural transition takes place at a smaller volume around the 13\% contraction.
The experimentally determined metal-insulator transition volume for $Nd$NiO$_3$ occurs at a rather smaller reduced volume (around $2\%$) than is found theoretically ($\sim 3.4\%$)%is overestimated than the experimental reduced volume of NdNiO$_3$ (around -2\%, the black dotted line).
 
$La$NiO$_3$ is the only nickelate in a rhombohedral structure experimentally and 
and at ambient pressure remains metallic without any bond-length disproportionation at down to lowest temperature.
%always metallic without any bond-length disproportionation at ambient pressure. 
The DFT+DMFT result predicts that the material would undergo a strongly first order transition to a bond disproportionated state at a slightly larger volume ($\sim1.5\%$).
This result is reminiscent of the metal-insulator transition of $La$NiO$_3$ observed in thin films under tensile strain~\cite{Son:10,Chakhal:11}.

\section{Electronic Entropy and the Thermally Driven Transition}

\begin{figure}[!htbp]
\includegraphics[scale=0.45]{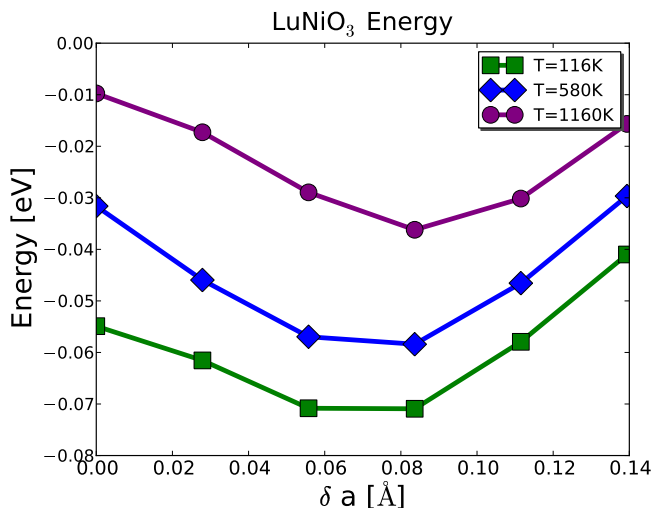}
\caption{(Color online)
Total energy curve in $Lu$NiO$_3$ as a function of the bond disproportionation $\delta a$  computed using DFT+DMFT at different temperatures ($T$=116K, 580K, 1160K). %The $P2_1/n$ strucutre is stable as the electronic ground state even above the experimental structural transition temperature in LuNiO$_3$ ($\sim$600K).At $T$=116K, the entropy contribution $T\cdot S$ term is estimated as 5meV lower for the Mott insulating $P2_1/n$ state compared to the metallic state at $\delta a=0$.
\label{fig:DMFT_T}}
\end{figure}

The nickelates undergo thermally driven insulator-metal and structure transitions with temperatures which can be as high as $\sim 500K$ for $Lu$NiO$_3$~\cite{Alonso:01}. Fig.$\:$\ref{fig:DMFT_T} depicts  the DFT+DMFT energy curve as a function of $\delta a$  computed for $Lu$NiO$_3$ at different temperatures.  We see that increasing the temperature does not change the electronic energetics significantly; both the location of the minimum and the magnitude of  the energy difference between undistorted and distorted states remain essentially unchanged even up to temperatures of more than twice the ordering temperature. We can also see that electronic entropy effects would only enhance the distorted state as the site selective Mott state which describes the insulating nickelates has a $S=1$ local moment on every second site corresponding to entropy $\mathbb{S} = k_B(\ln 3)/2 \simeq 0.55k_B$ per site ($k_B$ is the Boltzmann constant). The undistorted metallic state is a correlated Fermi liquid at low temperatures [$\mathbb{S} = (\pi^2/3)k_BD(e_f)T/Z \simeq 0.005k_B$  using the band theory values for the density of states $D$ and our calculated $Z\sim 3$ (consistent with other work~\cite{Deng:12})]. At higher temperatures the state evolves to a bad metal  with a very large scattering rate and the entropy increase saturates. Thus considerations of electronic entropy favor the distorted state at all accessible temperatures and we therefore conclude that the transition is driven by phonon entropy effects which are not included in our calculation.

\section{Conclusion}

In summary, we used state of the art DFT+DMFT methods to study theoretically the interplay of nontrivial structural and electronic effects in the rare earth   nickelates. We found that calculations using one fixed set of interaction parameters  correctly captures the dependence of structural   ($Pbnm$ vs $P2_1/n$) and electronic (metal vs insulator) properties as a function of rare earth ion and applied pressure (see Fig.$\:$\ref{fig:phasedmft}). Furthermore, the computed bond length disproportionation is in good agreement with experiment (see Fig.$\:$\ref{fig:bondvsvol}) while the thermal transition
has been shown to be driven by phonon entropy effects. These results establish the DFT+DMFT method as a useful tool for predicting structural and electronic properties of strongly correlated oxides. One important direction for future research is the computation of forces, which allow for first-principles
molecular dynamics based on DFT+DMFT energetics. 

$\textit{Acknowledgements:}$
The authors acknowledge funding from the U.S. Army Research Office via Grant
No. W911NF0910345 56032PH, US Department of Energy under Grant DOE-ER-046169,
and in part by FAME, one of six centers of STARnet, a
Semiconductor Research Corporation program sponsored by MARCO and DARPA.

\bibliography{main}

\begin{thebibliography}{41}%
\makeatletter
\providecommand \@ifxundefined [1]{%
 \@ifx{#1\undefined}
}%
\providecommand \@ifnum [1]{%
 \ifnum #1\expandafter \@firstoftwo
 \else \expandafter \@secondoftwo
 \fi
}%
\providecommand \@ifx [1]{%
 \ifx #1\expandafter \@firstoftwo
 \else \expandafter \@secondoftwo
 \fi
}%
\providecommand \natexlab [1]{#1}%
\providecommand \enquote  [1]{``#1''}%
\providecommand \bibnamefont  [1]{#1}%
\providecommand \bibfnamefont [1]{#1}%
\providecommand \citenamefont [1]{#1}%
\providecommand \href@noop [0]{\@secondoftwo}%
\providecommand \href [0]{\begingroup \@sanitize@url \@href}%
\providecommand \@href[1]{\@@startlink{#1}\@@href}%
\providecommand \@@href[1]{\endgroup#1\@@endlink}%
\providecommand \@sanitize@url [0]{\catcode `\\12\catcode `\$12\catcode
  `\&12\catcode `\#12\catcode `\^12\catcode `\_12\catcode `\%12\relax}%
\providecommand \@@startlink[1]{}%
\providecommand \@@endlink[0]{}%
\providecommand \url  [0]{\begingroup\@sanitize@url \@url }%
\providecommand \@url [1]{\endgroup\@href {#1}{\urlprefix }}%
\providecommand \urlprefix  [0]{URL }%
\providecommand \Eprint [0]{\href }%
\providecommand \doibase [0]{http://dx.doi.org/}%
\providecommand \selectlanguage [0]{\@gobble}%
\providecommand \bibinfo  [0]{\@secondoftwo}%
\providecommand \bibfield  [0]{\@secondoftwo}%
\providecommand \translation [1]{[#1]}%
\providecommand \BibitemOpen [0]{}%
\providecommand \bibitemStop [0]{}%
\providecommand \bibitemNoStop [0]{.\EOS\space}%
\providecommand \EOS [0]{\spacefactor3000\relax}%
\providecommand \BibitemShut  [1]{\csname bibitem#1\endcsname}%
\let\auto@bib@innerbib\@empty
%</preamble>
\bibitem [{\citenamefont {Kotliar}\ \emph {et~al.}(2006)\citenamefont
  {Kotliar}, \citenamefont {Savrasov}, \citenamefont {Haule}, \citenamefont
  {Oudovenko}, \citenamefont {Parcollet},\ and\ \citenamefont
  {Marianetti}}]{Kotliar:06}%
  \BibitemOpen
  \bibfield  {author} {\bibinfo {author} {\bibfnamefont {G.}~\bibnamefont
  {Kotliar}}, \bibinfo {author} {\bibfnamefont {S.~Y.}\ \bibnamefont
  {Savrasov}}, \bibinfo {author} {\bibfnamefont {K.}~\bibnamefont {Haule}},
  \bibinfo {author} {\bibfnamefont {V.~S.}\ \bibnamefont {Oudovenko}}, \bibinfo
  {author} {\bibfnamefont {O.}~\bibnamefont {Parcollet}}, \ and\ \bibinfo
  {author} {\bibfnamefont {C.~A.}\ \bibnamefont {Marianetti}},\ }\href
  {\doibase 10.1103/RevModPhys.78.865} {\bibfield  {journal} {\bibinfo
  {journal} {Rev. Mod. Phys.}\ }\textbf {\bibinfo {volume} {78}},\ \bibinfo
  {pages} {865} (\bibinfo {year} {2006})}\BibitemShut {NoStop}%
\bibitem [{\citenamefont {Savrasov}\ \emph {et~al.}(2001)\citenamefont
  {Savrasov}, \citenamefont {Kotliar},\ and\ \citenamefont
  {Abrahams}}]{Savrasov:01}%
  \BibitemOpen
  \bibfield  {author} {\bibinfo {author} {\bibfnamefont {S.~Y.}\ \bibnamefont
  {Savrasov}}, \bibinfo {author} {\bibfnamefont {G.}~\bibnamefont {Kotliar}}, \
  and\ \bibinfo {author} {\bibfnamefont {E.}~\bibnamefont {Abrahams}},\
  }\href@noop {} {\bibfield  {journal} {\bibinfo  {journal} {Nature}\ }\textbf
  {\bibinfo {volume} {410}},\ \bibinfo {pages} {793} (\bibinfo {year}
  {2001})}\BibitemShut {NoStop}%
\bibitem [{\citenamefont {Held}\ \emph {et~al.}(2001)\citenamefont {Held},
  \citenamefont {McMahan},\ and\ \citenamefont {Scalettar}}]{Held:01}%
  \BibitemOpen
  \bibfield  {author} {\bibinfo {author} {\bibfnamefont {K.}~\bibnamefont
  {Held}}, \bibinfo {author} {\bibfnamefont {A.~K.}\ \bibnamefont {McMahan}}, \
  and\ \bibinfo {author} {\bibfnamefont {R.~T.}\ \bibnamefont {Scalettar}},\
  }\href {\doibase 10.1103/PhysRevLett.87.276404} {\bibfield  {journal}
  {\bibinfo  {journal} {Phys. Rev. Lett.}\ }\textbf {\bibinfo {volume} {87}},\
  \bibinfo {pages} {276404} (\bibinfo {year} {2001})}\BibitemShut {NoStop}%
\bibitem [{\citenamefont {McMahan}\ \emph {et~al.}(2003)\citenamefont
  {McMahan}, \citenamefont {Held},\ and\ \citenamefont
  {Scalettar}}]{McMahan:03}%
  \BibitemOpen
  \bibfield  {author} {\bibinfo {author} {\bibfnamefont {A.~K.}\ \bibnamefont
  {McMahan}}, \bibinfo {author} {\bibfnamefont {K.}~\bibnamefont {Held}}, \
  and\ \bibinfo {author} {\bibfnamefont {R.~T.}\ \bibnamefont {Scalettar}},\
  }\href {\doibase 10.1103/PhysRevB.67.075108} {\bibfield  {journal} {\bibinfo
  {journal} {Phys. Rev. B}\ }\textbf {\bibinfo {volume} {67}},\ \bibinfo
  {pages} {075108} (\bibinfo {year} {2003})}\BibitemShut {NoStop}%
\bibitem [{\citenamefont {Amadon}\ \emph {et~al.}(2006)\citenamefont {Amadon},
  \citenamefont {Biermann}, \citenamefont {Georges},\ and\ \citenamefont
  {Aryasetiawan}}]{Amadon:06}%
  \BibitemOpen
  \bibfield  {author} {\bibinfo {author} {\bibfnamefont {B.}~\bibnamefont
  {Amadon}}, \bibinfo {author} {\bibfnamefont {S.}~\bibnamefont {Biermann}},
  \bibinfo {author} {\bibfnamefont {A.}~\bibnamefont {Georges}}, \ and\
  \bibinfo {author} {\bibfnamefont {F.}~\bibnamefont {Aryasetiawan}},\ }\href
  {\doibase 10.1103/PhysRevLett.96.066402} {\bibfield  {journal} {\bibinfo
  {journal} {Phys. Rev. Lett.}\ }\textbf {\bibinfo {volume} {96}},\ \bibinfo
  {pages} {066402} (\bibinfo {year} {2006})}\BibitemShut {NoStop}%
\bibitem [{\citenamefont {Di~Marco}\ \emph {et~al.}(2009)\citenamefont
  {Di~Marco}, \citenamefont {Min\'ar}, \citenamefont {Chadov}, \citenamefont
  {Katsnelson}, \citenamefont {Ebert},\ and\ \citenamefont
  {Lichtenstein}}]{Marco:09}%
  \BibitemOpen
  \bibfield  {author} {\bibinfo {author} {\bibfnamefont {I.}~\bibnamefont
  {Di~Marco}}, \bibinfo {author} {\bibfnamefont {J.}~\bibnamefont {Min\'ar}},
  \bibinfo {author} {\bibfnamefont {S.}~\bibnamefont {Chadov}}, \bibinfo
  {author} {\bibfnamefont {M.~I.}\ \bibnamefont {Katsnelson}}, \bibinfo
  {author} {\bibfnamefont {H.}~\bibnamefont {Ebert}}, \ and\ \bibinfo {author}
  {\bibfnamefont {A.~I.}\ \bibnamefont {Lichtenstein}},\ }\href {\doibase
  10.1103/PhysRevB.79.115111} {\bibfield  {journal} {\bibinfo  {journal} {Phys.
  Rev. B}\ }\textbf {\bibinfo {volume} {79}},\ \bibinfo {pages} {115111}
  (\bibinfo {year} {2009})}\BibitemShut {NoStop}%
\bibitem [{\citenamefont {Leonov}\ \emph {et~al.}(2008)\citenamefont {Leonov},
  \citenamefont {Binggeli}, \citenamefont {Korotin}, \citenamefont {Anisimov},
  \citenamefont {Stoji\ifmmode~\acute{c}\else \'{c}\fi{}},\ and\ \citenamefont
  {Vollhardt}}]{Leonov:08}%
  \BibitemOpen
  \bibfield  {author} {\bibinfo {author} {\bibfnamefont {I.}~\bibnamefont
  {Leonov}}, \bibinfo {author} {\bibfnamefont {N.}~\bibnamefont {Binggeli}},
  \bibinfo {author} {\bibfnamefont {D.}~\bibnamefont {Korotin}}, \bibinfo
  {author} {\bibfnamefont {V.~I.}\ \bibnamefont {Anisimov}}, \bibinfo {author}
  {\bibfnamefont {N.}~\bibnamefont {Stoji\ifmmode~\acute{c}\else \'{c}\fi{}}},
  \ and\ \bibinfo {author} {\bibfnamefont {D.}~\bibnamefont {Vollhardt}},\
  }\href {\doibase 10.1103/PhysRevLett.101.096405} {\bibfield  {journal}
  {\bibinfo  {journal} {Phys. Rev. Lett.}\ }\textbf {\bibinfo {volume} {101}},\
  \bibinfo {pages} {096405} (\bibinfo {year} {2008})}\BibitemShut {NoStop}%
\bibitem [{\citenamefont {Leonov}\ \emph {et~al.}(2010)\citenamefont {Leonov},
  \citenamefont {Korotin}, \citenamefont {Binggeli}, \citenamefont {Anisimov},\
  and\ \citenamefont {Vollhardt}}]{Leonov:10}%
  \BibitemOpen
  \bibfield  {author} {\bibinfo {author} {\bibfnamefont {I.}~\bibnamefont
  {Leonov}}, \bibinfo {author} {\bibfnamefont {D.}~\bibnamefont {Korotin}},
  \bibinfo {author} {\bibfnamefont {N.}~\bibnamefont {Binggeli}}, \bibinfo
  {author} {\bibfnamefont {V.~I.}\ \bibnamefont {Anisimov}}, \ and\ \bibinfo
  {author} {\bibfnamefont {D.}~\bibnamefont {Vollhardt}},\ }\href {\doibase
  10.1103/PhysRevB.81.075109} {\bibfield  {journal} {\bibinfo  {journal} {Phys.
  Rev. B}\ }\textbf {\bibinfo {volume} {81}},\ \bibinfo {pages} {075109}
  (\bibinfo {year} {2010})}\BibitemShut {NoStop}%
\bibitem [{\citenamefont {Leonov}\ \emph {et~al.}(2011)\citenamefont {Leonov},
  \citenamefont {Poteryaev}, \citenamefont {Anisimov},\ and\ \citenamefont
  {Vollhardt}}]{Leonov:11}%
  \BibitemOpen
  \bibfield  {author} {\bibinfo {author} {\bibfnamefont {I.}~\bibnamefont
  {Leonov}}, \bibinfo {author} {\bibfnamefont {A.~I.}\ \bibnamefont
  {Poteryaev}}, \bibinfo {author} {\bibfnamefont {V.~I.}\ \bibnamefont
  {Anisimov}}, \ and\ \bibinfo {author} {\bibfnamefont {D.}~\bibnamefont
  {Vollhardt}},\ }\href {\doibase 10.1103/PhysRevLett.106.106405} {\bibfield
  {journal} {\bibinfo  {journal} {Phys. Rev. Lett.}\ }\textbf {\bibinfo
  {volume} {106}},\ \bibinfo {pages} {106405} (\bibinfo {year}
  {2011})}\BibitemShut {NoStop}%
\bibitem [{\citenamefont {Amadon}(2012)}]{Amadon:12}%
  \BibitemOpen
  \bibfield  {author} {\bibinfo {author} {\bibfnamefont {B.}~\bibnamefont
  {Amadon}},\ }\href@noop {} {\bibfield  {journal} {\bibinfo  {journal}
  {Journal of Physics: Condensed Matter}\ }\textbf {\bibinfo {volume} {24}},\
  \bibinfo {pages} {075604} (\bibinfo {year} {2012})}\BibitemShut {NoStop}%
\bibitem [{\citenamefont {Pourovskii}\ \emph {et~al.}(2007)\citenamefont
  {Pourovskii}, \citenamefont {Amadon}, \citenamefont {Biermann},\ and\
  \citenamefont {Georges}}]{Pourovskii2007235101}%
  \BibitemOpen
  \bibfield  {author} {\bibinfo {author} {\bibfnamefont {L.~V.}\ \bibnamefont
  {Pourovskii}}, \bibinfo {author} {\bibfnamefont {B.}~\bibnamefont {Amadon}},
  \bibinfo {author} {\bibfnamefont {S.}~\bibnamefont {Biermann}}, \ and\
  \bibinfo {author} {\bibfnamefont {A.}~\bibnamefont {Georges}},\ }\href
  {\doibase 10.1103/PhysRevB.76.235101} {\bibfield  {journal} {\bibinfo
  {journal} {Phys. Rev. B}\ }\textbf {\bibinfo {volume} {76}},\ \bibinfo
  {pages} {235101} (\bibinfo {year} {2007})}\BibitemShut {NoStop}%
\bibitem [{\citenamefont {Aichhorn}\ \emph {et~al.}(2011)\citenamefont
  {Aichhorn}, \citenamefont {Pourovskii},\ and\ \citenamefont
  {Georges}}]{Aichhorn:11}%
  \BibitemOpen
  \bibfield  {author} {\bibinfo {author} {\bibfnamefont {M.}~\bibnamefont
  {Aichhorn}}, \bibinfo {author} {\bibfnamefont {L.}~\bibnamefont
  {Pourovskii}}, \ and\ \bibinfo {author} {\bibfnamefont {A.}~\bibnamefont
  {Georges}},\ }\href {\doibase 10.1103/PhysRevB.84.054529} {\bibfield
  {journal} {\bibinfo  {journal} {Phys. Rev. B}\ }\textbf {\bibinfo {volume}
  {84}},\ \bibinfo {pages} {054529} (\bibinfo {year} {2011})}\BibitemShut
  {NoStop}%
\bibitem [{\citenamefont {Lee}\ \emph {et~al.}(2012)\citenamefont {Lee},
  \citenamefont {Ji}, \citenamefont {Kim}, \citenamefont {Kim}, \citenamefont
  {Haule}, \citenamefont {Kotliar}, \citenamefont {Lee}, \citenamefont {Khim},
  \citenamefont {Kim}, \citenamefont {Kim}, \citenamefont {Kim},\ and\
  \citenamefont {Shim}}]{Lee:12}%
  \BibitemOpen
  \bibfield  {author} {\bibinfo {author} {\bibfnamefont {G.}~\bibnamefont
  {Lee}}, \bibinfo {author} {\bibfnamefont {H.~S.}\ \bibnamefont {Ji}},
  \bibinfo {author} {\bibfnamefont {Y.}~\bibnamefont {Kim}}, \bibinfo {author}
  {\bibfnamefont {C.}~\bibnamefont {Kim}}, \bibinfo {author} {\bibfnamefont
  {K.}~\bibnamefont {Haule}}, \bibinfo {author} {\bibfnamefont
  {G.}~\bibnamefont {Kotliar}}, \bibinfo {author} {\bibfnamefont
  {B.}~\bibnamefont {Lee}}, \bibinfo {author} {\bibfnamefont {S.}~\bibnamefont
  {Khim}}, \bibinfo {author} {\bibfnamefont {K.~H.}\ \bibnamefont {Kim}},
  \bibinfo {author} {\bibfnamefont {K.~S.}\ \bibnamefont {Kim}}, \bibinfo
  {author} {\bibfnamefont {K.-S.}\ \bibnamefont {Kim}}, \ and\ \bibinfo
  {author} {\bibfnamefont {J.~H.}\ \bibnamefont {Shim}},\ }\href {\doibase
  10.1103/PhysRevLett.109.177001} {\bibfield  {journal} {\bibinfo  {journal}
  {Phys. Rev. Lett.}\ }\textbf {\bibinfo {volume} {109}},\ \bibinfo {pages}
  {177001} (\bibinfo {year} {2012})}\BibitemShut {NoStop}%
\bibitem [{\citenamefont {Grieger}\ \emph {et~al.}(2012)\citenamefont
  {Grieger}, \citenamefont {Piefke}, \citenamefont {Peil},\ and\ \citenamefont
  {Lechermann}}]{Lechermann:12}%
  \BibitemOpen
  \bibfield  {author} {\bibinfo {author} {\bibfnamefont {D.}~\bibnamefont
  {Grieger}}, \bibinfo {author} {\bibfnamefont {C.}~\bibnamefont {Piefke}},
  \bibinfo {author} {\bibfnamefont {O.~E.}\ \bibnamefont {Peil}}, \ and\
  \bibinfo {author} {\bibfnamefont {F.}~\bibnamefont {Lechermann}},\
  }\href@noop {} {\bibfield  {journal} {\bibinfo  {journal} {Phys. Rev. B}\
  }\textbf {\bibinfo {volume} {86}},\ \bibinfo {pages} {155121} (\bibinfo
  {year} {2012})}\BibitemShut {NoStop}%
\bibitem [{\citenamefont {Bieder}\ and\ \citenamefont
  {Amadon}(2014)}]{Amadon:13}%
  \BibitemOpen
  \bibfield  {author} {\bibinfo {author} {\bibfnamefont {J.}~\bibnamefont
  {Bieder}}\ and\ \bibinfo {author} {\bibfnamefont {B.}~\bibnamefont
  {Amadon}},\ }\href {\doibase 10.1103/PhysRevB.89.195132} {\bibfield
  {journal} {\bibinfo  {journal} {Phys. Rev. B}\ }\textbf {\bibinfo {volume}
  {89}},\ \bibinfo {pages} {195132} (\bibinfo {year} {2014})}\BibitemShut
  {NoStop}%
\bibitem [{\citenamefont {Medarde}(1997)}]{Medarde:97}%
  \BibitemOpen
  \bibfield  {author} {\bibinfo {author} {\bibfnamefont {M.~L.}\ \bibnamefont
  {Medarde}},\ }\href@noop {} {\bibfield  {journal} {\bibinfo  {journal}
  {Journal of Physics: Condensed Matter}\ }\textbf {\bibinfo {volume} {9}},\
  \bibinfo {pages} {1679} (\bibinfo {year} {1997})}\BibitemShut {NoStop}%
\bibitem [{\citenamefont {Alonso}\ \emph {et~al.}(1999)\citenamefont {Alonso},
  \citenamefont {Garcia-Munoz}, \citenamefont {Fernandez-Diaz}, \citenamefont
  {Aranda}, \citenamefont {Martinez-Lope},\ and\ \citenamefont
  {Casais}}]{Alonso:99}%
  \BibitemOpen
  \bibfield  {author} {\bibinfo {author} {\bibfnamefont {J.~A.}\ \bibnamefont
  {Alonso}}, \bibinfo {author} {\bibfnamefont {J.~L.}\ \bibnamefont
  {Garcia-Munoz}}, \bibinfo {author} {\bibfnamefont {M.~T.}\ \bibnamefont
  {Fernandez-Diaz}}, \bibinfo {author} {\bibfnamefont {M.~A.~G.}\ \bibnamefont
  {Aranda}}, \bibinfo {author} {\bibfnamefont {M.~J.}\ \bibnamefont
  {Martinez-Lope}}, \ and\ \bibinfo {author} {\bibfnamefont {M.~T.}\
  \bibnamefont {Casais}},\ }\href {\doibase 10.1103/PhysRevLett.82.3871}
  {\bibfield  {journal} {\bibinfo  {journal} {Phys. Rev. Lett.}\ }\textbf
  {\bibinfo {volume} {82}},\ \bibinfo {pages} {3871} (\bibinfo {year}
  {1999})}\BibitemShut {NoStop}%
\bibitem [{\citenamefont {Alonso}\ \emph {et~al.}(2001)\citenamefont {Alonso},
  \citenamefont {Martinez-Lope}, \citenamefont {Casais}, \citenamefont
  {Garcia-Munoz}, \citenamefont {Fernandez-Diaz},\ and\ \citenamefont
  {Aranda}}]{Alonso:01}%
  \BibitemOpen
  \bibfield  {author} {\bibinfo {author} {\bibfnamefont {J.~A.}\ \bibnamefont
  {Alonso}}, \bibinfo {author} {\bibfnamefont {M.~J.}\ \bibnamefont
  {Martinez-Lope}}, \bibinfo {author} {\bibfnamefont {M.~T.}\ \bibnamefont
  {Casais}}, \bibinfo {author} {\bibfnamefont {J.~L.}\ \bibnamefont
  {Garcia-Munoz}}, \bibinfo {author} {\bibfnamefont {M.~T.}\ \bibnamefont
  {Fernandez-Diaz}}, \ and\ \bibinfo {author} {\bibfnamefont {M.~A.~G.}\
  \bibnamefont {Aranda}},\ }\href@noop {} {\bibfield  {journal} {\bibinfo
  {journal} {Phys. Rev. B}\ }\textbf {\bibinfo {volume} {64}},\ \bibinfo
  {pages} {094102} (\bibinfo {year} {2001})}\BibitemShut {NoStop}%
\bibitem [{\citenamefont {Staub}\ \emph {et~al.}(2002)\citenamefont {Staub},
  \citenamefont {Meijer}, \citenamefont {Fauth}, \citenamefont {Allenspach},
  \citenamefont {Bednorz}, \citenamefont {Karpinski}, \citenamefont {Kazakov},
  \citenamefont {Paolasini},\ and\ \citenamefont {d'Acapito}}]{Staub:02}%
  \BibitemOpen
  \bibfield  {author} {\bibinfo {author} {\bibfnamefont {U.}~\bibnamefont
  {Staub}}, \bibinfo {author} {\bibfnamefont {G.~I.}\ \bibnamefont {Meijer}},
  \bibinfo {author} {\bibfnamefont {F.}~\bibnamefont {Fauth}}, \bibinfo
  {author} {\bibfnamefont {R.}~\bibnamefont {Allenspach}}, \bibinfo {author}
  {\bibfnamefont {J.~G.}\ \bibnamefont {Bednorz}}, \bibinfo {author}
  {\bibfnamefont {J.}~\bibnamefont {Karpinski}}, \bibinfo {author}
  {\bibfnamefont {S.~M.}\ \bibnamefont {Kazakov}}, \bibinfo {author}
  {\bibfnamefont {L.}~\bibnamefont {Paolasini}}, \ and\ \bibinfo {author}
  {\bibfnamefont {F.}~\bibnamefont {d'Acapito}},\ }\href@noop {} {\bibfield
  {journal} {\bibinfo  {journal} {Phys. Rev. Lett.}\ }\textbf {\bibinfo
  {volume} {88}},\ \bibinfo {pages} {126402} (\bibinfo {year}
  {2002})}\BibitemShut {NoStop}%
\bibitem [{\citenamefont {Mazin}\ \emph {et~al.}(2007)\citenamefont {Mazin},
  \citenamefont {Khomskii}, \citenamefont {Lengsdorf}, \citenamefont {Alonso},
  \citenamefont {Marshall}, \citenamefont {Ibberson}, \citenamefont
  {Podlesnyak}, \citenamefont {Martinez-Lope},\ and\ \citenamefont
  {Abd-Elmeguid}}]{Mazin:07}%
  \BibitemOpen
  \bibfield  {author} {\bibinfo {author} {\bibfnamefont {I.~I.}\ \bibnamefont
  {Mazin}}, \bibinfo {author} {\bibfnamefont {D.~I.}\ \bibnamefont {Khomskii}},
  \bibinfo {author} {\bibfnamefont {R.}~\bibnamefont {Lengsdorf}}, \bibinfo
  {author} {\bibfnamefont {J.~A.}\ \bibnamefont {Alonso}}, \bibinfo {author}
  {\bibfnamefont {W.~G.}\ \bibnamefont {Marshall}}, \bibinfo {author}
  {\bibfnamefont {R.~M.}\ \bibnamefont {Ibberson}}, \bibinfo {author}
  {\bibfnamefont {A.}~\bibnamefont {Podlesnyak}}, \bibinfo {author}
  {\bibfnamefont {M.~J.}\ \bibnamefont {Martinez-Lope}}, \ and\ \bibinfo
  {author} {\bibfnamefont {M.~M.}\ \bibnamefont {Abd-Elmeguid}},\ }\href@noop
  {} {\bibfield  {journal} {\bibinfo  {journal} {Phys. Rev. Lett.}\ }\textbf
  {\bibinfo {volume} {98}},\ \bibinfo {pages} {176406} (\bibinfo {year}
  {2007})}\BibitemShut {NoStop}%
\bibitem [{\citenamefont {Medarde}\ \emph {et~al.}(2009)\citenamefont
  {Medarde}, \citenamefont {Dallera}, \citenamefont {Grioni}, \citenamefont
  {Delley}, \citenamefont {Vernay}, \citenamefont {Mesot}, \citenamefont
  {Sikora}, \citenamefont {Alonso},\ and\ \citenamefont
  {Martinez-Lope}}]{Medarde:09}%
  \BibitemOpen
  \bibfield  {author} {\bibinfo {author} {\bibfnamefont {M.}~\bibnamefont
  {Medarde}}, \bibinfo {author} {\bibfnamefont {C.}~\bibnamefont {Dallera}},
  \bibinfo {author} {\bibfnamefont {M.}~\bibnamefont {Grioni}}, \bibinfo
  {author} {\bibfnamefont {B.}~\bibnamefont {Delley}}, \bibinfo {author}
  {\bibfnamefont {F.}~\bibnamefont {Vernay}}, \bibinfo {author} {\bibfnamefont
  {J.}~\bibnamefont {Mesot}}, \bibinfo {author} {\bibfnamefont
  {M.}~\bibnamefont {Sikora}}, \bibinfo {author} {\bibfnamefont {J.~A.}\
  \bibnamefont {Alonso}}, \ and\ \bibinfo {author} {\bibfnamefont {M.~J.}\
  \bibnamefont {Martinez-Lope}},\ }\href@noop {} {\bibfield  {journal}
  {\bibinfo  {journal} {Phys. Rev. B}\ }\textbf {\bibinfo {volume} {80}},\
  \bibinfo {pages} {245105} (\bibinfo {year} {2009})}\BibitemShut {NoStop}%
\bibitem [{\citenamefont {Lee}\ \emph {et~al.}(2011)\citenamefont {Lee},
  \citenamefont {Chen},\ and\ \citenamefont {Balents}}]{Lee:11}%
  \BibitemOpen
  \bibfield  {author} {\bibinfo {author} {\bibfnamefont {S.~B.}\ \bibnamefont
  {Lee}}, \bibinfo {author} {\bibfnamefont {R.}~\bibnamefont {Chen}}, \ and\
  \bibinfo {author} {\bibfnamefont {L.}~\bibnamefont {Balents}},\ }\href@noop
  {} {\bibfield  {journal} {\bibinfo  {journal} {Phys. Rev. B}\ }\textbf
  {\bibinfo {volume} {84}},\ \bibinfo {pages} {165119} (\bibinfo {year}
  {2011})}\BibitemShut {NoStop}%
\bibitem [{\citenamefont {Park}\ \emph {et~al.}(2012)\citenamefont {Park},
  \citenamefont {Millis},\ and\ \citenamefont {Marianetti}}]{Park:12}%
  \BibitemOpen
  \bibfield  {author} {\bibinfo {author} {\bibfnamefont {H.}~\bibnamefont
  {Park}}, \bibinfo {author} {\bibfnamefont {A.~J.}\ \bibnamefont {Millis}}, \
  and\ \bibinfo {author} {\bibfnamefont {C.~A.}\ \bibnamefont {Marianetti}},\
  }\href {\doibase 10.1103/PhysRevLett.109.156402} {\bibfield  {journal}
  {\bibinfo  {journal} {Phys. Rev. Lett.}\ }\textbf {\bibinfo {volume} {109}},\
  \bibinfo {pages} {156402} (\bibinfo {year} {2012})}\BibitemShut {NoStop}%
\bibitem [{\citenamefont {Lengsdorf}\ \emph {et~al.}(2004)\citenamefont
  {Lengsdorf}, \citenamefont {Barla}, \citenamefont {Alonso}, \citenamefont
  {Martinez-Lope}, \citenamefont {Micklitz},\ and\ \citenamefont
  {Abd-Elmeguid}}]{Lengsdorf:04}%
  \BibitemOpen
  \bibfield  {author} {\bibinfo {author} {\bibfnamefont {R.}~\bibnamefont
  {Lengsdorf}}, \bibinfo {author} {\bibfnamefont {A.}~\bibnamefont {Barla}},
  \bibinfo {author} {\bibfnamefont {J.~A.}\ \bibnamefont {Alonso}}, \bibinfo
  {author} {\bibfnamefont {M.~J.}\ \bibnamefont {Martinez-Lope}}, \bibinfo
  {author} {\bibfnamefont {H.}~\bibnamefont {Micklitz}}, \ and\ \bibinfo
  {author} {\bibfnamefont {M.~M.}\ \bibnamefont {Abd-Elmeguid}},\ }\href@noop
  {} {\bibfield  {journal} {\bibinfo  {journal} {Journal of Physics: Condensed
  Matter}\ }\textbf {\bibinfo {volume} {16}},\ \bibinfo {pages} {3355}
  (\bibinfo {year} {2004})}\BibitemShut {NoStop}%
\bibitem [{\citenamefont {Garcia-Mu\~noz}\ \emph {et~al.}(2004)\citenamefont
  {Garcia-Mu\~noz}, \citenamefont {Amboage}, \citenamefont {Hanfland},
  \citenamefont {Alonso}, \citenamefont {Martinez-Lope},\ and\ \citenamefont
  {Mortimer}}]{Amboage:04}%
  \BibitemOpen
  \bibfield  {author} {\bibinfo {author} {\bibfnamefont {J.~L.}\ \bibnamefont
  {Garcia-Mu\~noz}}, \bibinfo {author} {\bibfnamefont {M.}~\bibnamefont
  {Amboage}}, \bibinfo {author} {\bibfnamefont {M.}~\bibnamefont {Hanfland}},
  \bibinfo {author} {\bibfnamefont {J.~A.}\ \bibnamefont {Alonso}}, \bibinfo
  {author} {\bibfnamefont {M.~J.}\ \bibnamefont {Martinez-Lope}}, \ and\
  \bibinfo {author} {\bibfnamefont {R.}~\bibnamefont {Mortimer}},\ }\href
  {\doibase 10.1103/PhysRevB.69.094106} {\bibfield  {journal} {\bibinfo
  {journal} {Phys. Rev. B}\ }\textbf {\bibinfo {volume} {69}},\ \bibinfo
  {pages} {094106} (\bibinfo {year} {2004})}\BibitemShut {NoStop}%
\bibitem [{\citenamefont {Amboage}\ \emph {et~al.}(2005)\citenamefont
  {Amboage}, \citenamefont {Hanfland}, \citenamefont {Alonso},\ and\
  \citenamefont {Martínez-Lope}}]{Amboage:05}%
  \BibitemOpen
  \bibfield  {author} {\bibinfo {author} {\bibfnamefont {M.}~\bibnamefont
  {Amboage}}, \bibinfo {author} {\bibfnamefont {M.}~\bibnamefont {Hanfland}},
  \bibinfo {author} {\bibfnamefont {J.~A.}\ \bibnamefont {Alonso}}, \ and\
  \bibinfo {author} {\bibfnamefont {M.~J.}\ \bibnamefont {Martínez-Lope}},\
  }\href@noop {} {\bibfield  {journal} {\bibinfo  {journal} {Journal of
  Physics: Condensed Matter}\ }\textbf {\bibinfo {volume} {17}},\ \bibinfo
  {pages} {S783} (\bibinfo {year} {2005})}\BibitemShut {NoStop}%
\bibitem [{\citenamefont {Cheng}\ \emph {et~al.}(2010)\citenamefont {Cheng},
  \citenamefont {Zhou}, \citenamefont {Goodenough}, \citenamefont {Alonso},\
  and\ \citenamefont {Martinez-Lope}}]{Cheng:10}%
  \BibitemOpen
  \bibfield  {author} {\bibinfo {author} {\bibfnamefont {J.-G.}\ \bibnamefont
  {Cheng}}, \bibinfo {author} {\bibfnamefont {J.-S.}\ \bibnamefont {Zhou}},
  \bibinfo {author} {\bibfnamefont {J.~B.}\ \bibnamefont {Goodenough}},
  \bibinfo {author} {\bibfnamefont {J.~A.}\ \bibnamefont {Alonso}}, \ and\
  \bibinfo {author} {\bibfnamefont {M.~J.}\ \bibnamefont {Martinez-Lope}},\
  }\href {\doibase 10.1103/PhysRevB.82.085107} {\bibfield  {journal} {\bibinfo
  {journal} {Phys. Rev. B}\ }\textbf {\bibinfo {volume} {82}},\ \bibinfo
  {pages} {085107} (\bibinfo {year} {2010})}\BibitemShut {NoStop}%
\bibitem [{\citenamefont {Anisimov}\ \emph {et~al.}(1991)\citenamefont
  {Anisimov}, \citenamefont {Zaanen},\ and\ \citenamefont
  {Andersen}}]{Anisimov91}%
  \BibitemOpen
  \bibfield  {author} {\bibinfo {author} {\bibfnamefont {V.~I.}\ \bibnamefont
  {Anisimov}}, \bibinfo {author} {\bibfnamefont {J.}~\bibnamefont {Zaanen}}, \
  and\ \bibinfo {author} {\bibfnamefont {O.~K.}\ \bibnamefont {Andersen}},\
  }\href {\doibase 10.1103/PhysRevB.44.943} {\bibfield  {journal} {\bibinfo
  {journal} {Phys. Rev. B}\ }\textbf {\bibinfo {volume} {44}},\ \bibinfo
  {pages} {943} (\bibinfo {year} {1991})}\BibitemShut {NoStop}%
\bibitem [{\citenamefont {Czyzyk}\ and\ \citenamefont
  {Sawatzky}(1994)}]{Sawatzky:94}%
  \BibitemOpen
  \bibfield  {author} {\bibinfo {author} {\bibfnamefont {M.~T.}\ \bibnamefont
  {Czyzyk}}\ and\ \bibinfo {author} {\bibfnamefont {G.~A.}\ \bibnamefont
  {Sawatzky}},\ }\href {\doibase 10.1103/PhysRevB.49.14211} {\bibfield
  {journal} {\bibinfo  {journal} {Phys. Rev. B}\ }\textbf {\bibinfo {volume}
  {49}},\ \bibinfo {pages} {14211} (\bibinfo {year} {1994})}\BibitemShut
  {NoStop}%
\bibitem [{\citenamefont {Amadon}\ \emph {et~al.}(2008)\citenamefont {Amadon},
  \citenamefont {Lechermann}, \citenamefont {Georges}, \citenamefont {Jollet},
  \citenamefont {Wehling},\ and\ \citenamefont {Lichtenstein}}]{Amadon08}%
  \BibitemOpen
  \bibfield  {author} {\bibinfo {author} {\bibfnamefont {B.}~\bibnamefont
  {Amadon}}, \bibinfo {author} {\bibfnamefont {F.}~\bibnamefont {Lechermann}},
  \bibinfo {author} {\bibfnamefont {A.}~\bibnamefont {Georges}}, \bibinfo
  {author} {\bibfnamefont {F.}~\bibnamefont {Jollet}}, \bibinfo {author}
  {\bibfnamefont {T.~O.}\ \bibnamefont {Wehling}}, \ and\ \bibinfo {author}
  {\bibfnamefont {A.~I.}\ \bibnamefont {Lichtenstein}},\ }\href {\doibase
  10.1103/PhysRevB.77.205112} {\bibfield  {journal} {\bibinfo  {journal} {Phys.
  Rev. B}\ }\textbf {\bibinfo {volume} {77}},\ \bibinfo {pages} {205112}
  (\bibinfo {year} {2008})}\BibitemShut {NoStop}%
\bibitem [{\citenamefont {Karolak}\ \emph {et~al.}(2010)\citenamefont
  {Karolak}, \citenamefont {Ulm}, \citenamefont {Wehling}, \citenamefont
  {Mazurenko}, \citenamefont {Poteryaev},\ and\ \citenamefont
  {Lichtenstein}}]{Karolak10}%
  \BibitemOpen
  \bibfield  {author} {\bibinfo {author} {\bibfnamefont {M.}~\bibnamefont
  {Karolak}}, \bibinfo {author} {\bibfnamefont {G.}~\bibnamefont {Ulm}},
  \bibinfo {author} {\bibfnamefont {T.}~\bibnamefont {Wehling}}, \bibinfo
  {author} {\bibfnamefont {V.}~\bibnamefont {Mazurenko}}, \bibinfo {author}
  {\bibfnamefont {A.}~\bibnamefont {Poteryaev}}, \ and\ \bibinfo {author}
  {\bibfnamefont {A.}~\bibnamefont {Lichtenstein}},\ }\href@noop {} {\bibfield
  {journal} {\bibinfo  {journal} {Journal of Electron Spectroscopy and Related
  Phenomena}\ }\textbf {\bibinfo {volume} {181}},\ \bibinfo {pages} {11}
  (\bibinfo {year} {2010})}\BibitemShut {NoStop}%
\bibitem [{\citenamefont {Wang}\ \emph {et~al.}(2012)\citenamefont {Wang},
  \citenamefont {Han}, \citenamefont {de' Medici}, \citenamefont {Park},
  \citenamefont {Marianetti},\ and\ \citenamefont {Millis}}]{Wang:12}%
  \BibitemOpen
  \bibfield  {author} {\bibinfo {author} {\bibfnamefont {X.}~\bibnamefont
  {Wang}}, \bibinfo {author} {\bibfnamefont {M.~J.}\ \bibnamefont {Han}},
  \bibinfo {author} {\bibfnamefont {L.}~\bibnamefont {de' Medici}}, \bibinfo
  {author} {\bibfnamefont {H.}~\bibnamefont {Park}}, \bibinfo {author}
  {\bibfnamefont {C.~A.}\ \bibnamefont {Marianetti}}, \ and\ \bibinfo {author}
  {\bibfnamefont {A.~J.}\ \bibnamefont {Millis}},\ }\href {\doibase
  10.1103/PhysRevB.86.195136} {\bibfield  {journal} {\bibinfo  {journal} {Phys.
  Rev. B}\ }\textbf {\bibinfo {volume} {86}},\ \bibinfo {pages} {195136}
  (\bibinfo {year} {2012})}\BibitemShut {NoStop}%
\bibitem [{\citenamefont {Dang}\ \emph {et~al.}(2014)\citenamefont {Dang},
  \citenamefont {Millis},\ and\ \citenamefont {Marianetti}}]{Dang:13}%
  \BibitemOpen
  \bibfield  {author} {\bibinfo {author} {\bibfnamefont {H.~T.}\ \bibnamefont
  {Dang}}, \bibinfo {author} {\bibfnamefont {A.~J.}\ \bibnamefont {Millis}}, \
  and\ \bibinfo {author} {\bibfnamefont {C.~A.}\ \bibnamefont {Marianetti}},\
  }\href {\doibase 10.1103/PhysRevB.89.161113} {\bibfield  {journal} {\bibinfo
  {journal} {Phys. Rev. B}\ }\textbf {\bibinfo {volume} {89}},\ \bibinfo
  {pages} {161113} (\bibinfo {year} {2014})}\BibitemShut {NoStop}%
\bibitem [{\citenamefont {Kresse}\ and\ \citenamefont
  {Joubert}(1999)}]{Kresse19991758}%
  \BibitemOpen
  \bibfield  {author} {\bibinfo {author} {\bibfnamefont {G.}~\bibnamefont
  {Kresse}}\ and\ \bibinfo {author} {\bibfnamefont {D.}~\bibnamefont
  {Joubert}},\ }\href@noop {} {\bibfield  {journal} {\bibinfo  {journal} {Phys.
  Rev. B}\ }\textbf {\bibinfo {volume} {59}},\ \bibinfo {pages} {1758}
  (\bibinfo {year} {1999})}\BibitemShut {NoStop}%
\bibitem [{\citenamefont {Kresse}\ and\ \citenamefont
  {Furthmuller}(1996)}]{Kresse199611169}%
  \BibitemOpen
  \bibfield  {author} {\bibinfo {author} {\bibfnamefont {G.}~\bibnamefont
  {Kresse}}\ and\ \bibinfo {author} {\bibfnamefont {J.}~\bibnamefont
  {Furthmuller}},\ }\href@noop {} {\bibfield  {journal} {\bibinfo  {journal}
  {Phys. Rev. B}\ }\textbf {\bibinfo {volume} {54}},\ \bibinfo {pages} {11169}
  (\bibinfo {year} {1996})}\BibitemShut {NoStop}%
\bibitem [{\citenamefont {Mostofi}\ \emph {et~al.}(2008)\citenamefont
  {Mostofi}, \citenamefont {Yates}, \citenamefont {Lee}, \citenamefont {Souza},
  \citenamefont {Vanderbilt},\ and\ \citenamefont {Marzari}}]{Wannier}%
  \BibitemOpen
  \bibfield  {author} {\bibinfo {author} {\bibfnamefont {A.~A.}\ \bibnamefont
  {Mostofi}}, \bibinfo {author} {\bibfnamefont {J.~R.}\ \bibnamefont {Yates}},
  \bibinfo {author} {\bibfnamefont {Y.-S.}\ \bibnamefont {Lee}}, \bibinfo
  {author} {\bibfnamefont {I.}~\bibnamefont {Souza}}, \bibinfo {author}
  {\bibfnamefont {D.}~\bibnamefont {Vanderbilt}}, \ and\ \bibinfo {author}
  {\bibfnamefont {N.}~\bibnamefont {Marzari}},\ }\href@noop {} {\bibfield
  {journal} {\bibinfo  {journal} {Computer Physics Communications}\ }\textbf
  {\bibinfo {volume} {178}},\ \bibinfo {pages} {685} (\bibinfo {year}
  {2008})}\BibitemShut {NoStop}%
\bibitem [{\citenamefont {Werner}\ \emph {et~al.}(2006)\citenamefont {Werner},
  \citenamefont {Comanac}, \citenamefont {deMedici}, \citenamefont {Troyer},\
  and\ \citenamefont {Millis}}]{Werner:06}%
  \BibitemOpen
  \bibfield  {author} {\bibinfo {author} {\bibfnamefont {P.}~\bibnamefont
  {Werner}}, \bibinfo {author} {\bibfnamefont {A.}~\bibnamefont {Comanac}},
  \bibinfo {author} {\bibfnamefont {L.}~\bibnamefont {deMedici}}, \bibinfo
  {author} {\bibfnamefont {M.}~\bibnamefont {Troyer}}, \ and\ \bibinfo {author}
  {\bibfnamefont {A.~J.}\ \bibnamefont {Millis}},\ }\href@noop {} {\bibfield
  {journal} {\bibinfo  {journal} {Phys. Rev. Lett.}\ }\textbf {\bibinfo
  {volume} {97}},\ \bibinfo {pages} {076405} (\bibinfo {year}
  {2006})}\BibitemShut {NoStop}%
\bibitem [{\citenamefont {Haule}(2007)}]{Haule:07}%
  \BibitemOpen
  \bibfield  {author} {\bibinfo {author} {\bibfnamefont {K.}~\bibnamefont
  {Haule}},\ }\href {\doibase 10.1103/PhysRevB.75.155113} {\bibfield  {journal}
  {\bibinfo  {journal} {Phys. Rev. B}\ }\textbf {\bibinfo {volume} {75}},\
  \bibinfo {pages} {155113} (\bibinfo {year} {2007})}\BibitemShut {NoStop}%
\bibitem [{\citenamefont {Son}\ \emph {et~al.}(2010)\citenamefont {Son},
  \citenamefont {Moetakef}, \citenamefont {Lebeau}, \citenamefont {Ouellette},
  \citenamefont {Balents}, \citenamefont {Allen},\ and\ \citenamefont
  {Stemmer}}]{Son:10}%
  \BibitemOpen
  \bibfield  {author} {\bibinfo {author} {\bibfnamefont {J.}~\bibnamefont
  {Son}}, \bibinfo {author} {\bibfnamefont {P.}~\bibnamefont {Moetakef}},
  \bibinfo {author} {\bibfnamefont {J.~M.}\ \bibnamefont {Lebeau}}, \bibinfo
  {author} {\bibfnamefont {D.}~\bibnamefont {Ouellette}}, \bibinfo {author}
  {\bibfnamefont {L.}~\bibnamefont {Balents}}, \bibinfo {author} {\bibfnamefont
  {S.~J.}\ \bibnamefont {Allen}}, \ and\ \bibinfo {author} {\bibfnamefont
  {S.}~\bibnamefont {Stemmer}},\ }\href@noop {} {\bibfield  {journal} {\bibinfo
   {journal} {Applied Physics Letters}\ }\textbf {\bibinfo {volume} {96}},\
  \bibinfo {pages} {062114} (\bibinfo {year} {2010})}\BibitemShut {NoStop}%
\bibitem [{\citenamefont {Chakhalian}\ \emph {et~al.}(2011)\citenamefont
  {Chakhalian}, \citenamefont {Rondinelli}, \citenamefont {Liu}, \citenamefont
  {Gray}, \citenamefont {Kareev}, \citenamefont {Moon}, \citenamefont {Prasai},
  \citenamefont {Cohn}, \citenamefont {Varela}, \citenamefont {Tung},
  \citenamefont {Bedzyk}, \citenamefont {Altendorf}, \citenamefont {Strigari},
  \citenamefont {Dabrowski}, \citenamefont {Tjeng}, \citenamefont {Ryan},\ and\
  \citenamefont {Freeland}}]{Chakhal:11}%
  \BibitemOpen
  \bibfield  {author} {\bibinfo {author} {\bibfnamefont {J.}~\bibnamefont
  {Chakhalian}}, \bibinfo {author} {\bibfnamefont {J.~M.}\ \bibnamefont
  {Rondinelli}}, \bibinfo {author} {\bibfnamefont {J.}~\bibnamefont {Liu}},
  \bibinfo {author} {\bibfnamefont {B.~A.}\ \bibnamefont {Gray}}, \bibinfo
  {author} {\bibfnamefont {M.}~\bibnamefont {Kareev}}, \bibinfo {author}
  {\bibfnamefont {E.~J.}\ \bibnamefont {Moon}}, \bibinfo {author}
  {\bibfnamefont {N.}~\bibnamefont {Prasai}}, \bibinfo {author} {\bibfnamefont
  {J.~L.}\ \bibnamefont {Cohn}}, \bibinfo {author} {\bibfnamefont
  {M.}~\bibnamefont {Varela}}, \bibinfo {author} {\bibfnamefont {I.~C.}\
  \bibnamefont {Tung}}, \bibinfo {author} {\bibfnamefont {M.~J.}\ \bibnamefont
  {Bedzyk}}, \bibinfo {author} {\bibfnamefont {S.~G.}\ \bibnamefont
  {Altendorf}}, \bibinfo {author} {\bibfnamefont {F.}~\bibnamefont {Strigari}},
  \bibinfo {author} {\bibfnamefont {B.}~\bibnamefont {Dabrowski}}, \bibinfo
  {author} {\bibfnamefont {L.~H.}\ \bibnamefont {Tjeng}}, \bibinfo {author}
  {\bibfnamefont {P.~J.}\ \bibnamefont {Ryan}}, \ and\ \bibinfo {author}
  {\bibfnamefont {J.~W.}\ \bibnamefont {Freeland}},\ }\href@noop {} {\bibfield
  {journal} {\bibinfo  {journal} {Phys. Rev. Lett.}\ }\textbf {\bibinfo
  {volume} {107}},\ \bibinfo {pages} {116805} (\bibinfo {year}
  {2011})}\BibitemShut {NoStop}%
\bibitem [{\citenamefont {Deng}\ \emph {et~al.}(2012)\citenamefont {Deng},
  \citenamefont {Ferrero}, \citenamefont {Mravlje}, \citenamefont {Aichhorn},\
  and\ \citenamefont {Georges}}]{Deng:12}%
  \BibitemOpen
  \bibfield  {author} {\bibinfo {author} {\bibfnamefont {X.}~\bibnamefont
  {Deng}}, \bibinfo {author} {\bibfnamefont {M.}~\bibnamefont {Ferrero}},
  \bibinfo {author} {\bibfnamefont {J.}~\bibnamefont {Mravlje}}, \bibinfo
  {author} {\bibfnamefont {M.}~\bibnamefont {Aichhorn}}, \ and\ \bibinfo
  {author} {\bibfnamefont {A.}~\bibnamefont {Georges}},\ }\href {\doibase
  10.1103/PhysRevB.85.125137} {\bibfield  {journal} {\bibinfo  {journal} {Phys.
  Rev. B}\ }\textbf {\bibinfo {volume} {85}},\ \bibinfo {pages} {125137}
  (\bibinfo {year} {2012})}\BibitemShut {NoStop}%
\end{thebibliography}%

\end{document}